\documentclass[%
a4paper,
 amsmath,amssymb,
 aps,prl,twocolumn,showpacs
]{revtex4}

\usepackage{graphicx}
\usepackage{dcolumn}
\usepackage{bm}
\usepackage{hyperref}

\begin{document}

\preprint{APS/123-QED}

\title{All-optical four-state magnetization reversal in (Ga,Mn)As ferromagnetic semiconductors}

\author{M. D. Kapetanakis$^{1}$}
\author{P. C. Lingos$^{1}$} 
\author{C. Piermarocchi$^{2}$}
\author{J. Wang$^{3}$}
\author{I. E. Perakis$^{1}$}
\affiliation{$^{1}$Department of Physics, University of Crete and
Institute of Electronic Structure \& Laser, Foundation for
Research and Technology-Hellas, Heraklion, Crete, 71110, Greece}
\affiliation{$^{2}$Department of Physics 
\& Astronomy,
 Michigan State University, East Lansing, MI, 48824, USA} 
\affiliation{$^{3}$Ames Laboratory and Department of Physics and Astronomy, Iowa State
University, Ames, Iowa 50011, USA}

\date{\today}

\begin{abstract}
Using density matrix equations of motion and  a tight--binding 
band calculation, we predict 
all--optical   switching 
between four metastable magnetic states  
of (III,Mn)As ferromagnets. 
This switching  
is initiated non--thermally within 100fs,  
during  nonlinear coherent photoexcitation.   
For a single optical pulse,  magnetization reversal is completed 
after $\sim$100 ps and controlled by the coherent femtosecond 
photoexcitation. 
Our predicted switching comes from 
magnetic nonlinearities
triggered by a femtosecond magnetization 
tilt that is  sensitive to un--adiabatic light--induced spin interactions.  
\end{abstract}

\pacs{78.47.J-, 78.20.Ls, 42.65.Re}

\maketitle

The goal of  THz magnetic switches 
underlies the entire field of
spin--electronics and 
challenges 
our understanding of  fundamental 
non--equilibrium spin  processes. 
The reading and writing of bits 
rely on reversing  the magnetization direction
between ``up'' and ``down''. 
In conventional switching,   
the magnetization moves out of  equilibrium  
via 
laser heating.  A  magnetic 
field then exerts a torque that reverses the 
magnetization within  few ns
\cite{stohr,schu}. 
This speed can be improved
by using 
coherent
spin rotation, via precession of the entire memory cell  
around a  magnetic field pulse-- precessional/ballistic switching \cite{schu,gerrits,stohr,tudosa}.
This  pulsed field  must have duration of at least half the precession period 
(100's of ps), which sets a fundamental limit of the magnetization reversal  time. 
This speed is further 
limited by 
randomness \cite{tudosa} or weak 
 precession damping 
allowing back-switching
of magnetic elements 
(``ringing'')
\cite{schu,gerrits}.
Faster switching, within  100fs,
could be explored by 
using  laser pulses to inject 
spin--polarized carriers \cite{kimel-rev}.
This is important for meeting the demand for 
 improved   read/write speeds, bit density,
and reliability
of current   magnetic devices.  

Magnetic properties of (III,Mn)V ferromagentic 
semiconductors exhibit sensitive response 
to carrier density tuning via light, 
electrical gates, or spin currents. This holds promise 
for high-speed magnetic switches
that combine information processing and storage 
on a single chip device 
with low power consumption
\cite{jung}.
The femtosecond photoexcitation of GaMnAs revealed 
distinct transient magneto-optical responses:
 (i) ultrafast decrease of the magnetization 
{\em amplitude} \cite{bigot1} within $\sim$ 100fs (demagnetization)
\cite{wang-rev,wang-demag,theory},
(ii) enhancement 
of magnetic order on ps timescale \cite{remag}, 
(iii) magnetization re--orientation within $\sim$100fs, 
followed by a distinct ps regime of coherent precession
\cite{kapet,wang-apl}.
Such non--equilibrium 
magnetic effects appear to be universal \cite{bigot1,bigot2,bigot}.
The pioneering work of Bigot and collaborators 
\cite{bigot1}, who observed 
demagnetization on a $\sim$100fs timescale
much shorter than the spin--phonon relaxation time,
 thus evolved into a new field of {\em femto--magnetism}. 
However, the  many--body theory of femto--magnetism  
 remains controversial \cite{bigot,kimel-rev}
and must ultimately engage  the 
elements of transient coherence,
 correlation, and nonlinearity
{\em on an equal footing}.
Here we present such a mean field theory
and propose a nonthermal mechanism  
for achieving ultrafast all--optical
magnetization switching in ferromagnetic (Ga,Mn)As.

We use a {\em single} 
100fs optical pulse
at $\sim$3eV,
 in resonance with the strong peak in 
the density of states for 
$\Lambda_3$$\rightarrow$$\Lambda_1$ 
interband transitions 
along the eight equivalent
  $\{111\}$ directions
 of the GaAs Brillouin zone 
(the $\Lambda$--edge)
\cite{burch}.
Magnetization switching is initiated 
within $\sim$100fs 
via {\em non--thermal} spin manipulation.
To describe this, we 
consider the  Hamiltonian 
$H(t)$=$H_{b}$+$H_{exch}(t)$+$H_{L}(t)$, 
where $H_b$=$H_{0}+H_{SO}+H_{pd}$ is the standard Hamiltonian 
describing  {\em un--excited} (Ga,Mn)As bands \cite{jung,vogl}.  
$H_{0}$ describes the states in the presence  of  the  lattice
 potential, while 
$H_{SO}$ is the spin-orbit interaction
\cite{vogl}. 
We consider hole densities $\sim$10$^{20}$cm$^{-3}$ 
 where 
the  virtual crystal approximation
applies 
\cite{jung}. 
 $H_{pd}$ is the mean--field 
interaction  \cite{jung,chovan} of the 
hole spin with the  {\em ground state} Mn spin 
${\bf S}_{0}$, 
whose  strength $\beta$=24 meV nm$^3$ 
was extracted from experiment \cite{wang-apl}. 
 $H_{pd}$
lifts 
the  degeneracy of the  GaAs  bands \cite{vogl} 
by the magnetic exchange energy
$\Delta_{pd}$=88meV.
Below  we describe microscopically a photo--induced magnetic anisotropy
characterized by the energy ratio 
$E_{SO}/\Delta_{pd}$ 
($E_{SO}$$\sim$350meV is the 
spin--orbit energy),  
which is {\em not} due to 
thermal \cite{bigot2} effects.
For this we start by 
diagonalizing $H_b$ using the Slater-Koster $sp^3s^*$ tight--binding
Hamiltonian \cite{vogl}. We thus  
obtain a basis of valence hole (conduction electron) states  
created by $\hat{h}_{{\bf k} n}^{\dag}$ ($\hat{e}_{{\bf k}n}^{\dag}$), where   
${\bf k}$ is the crystal momentum and  
$n$ labels the 20  bands \cite{vogl}.
This static bandstructure is modified  
by the interaction 
$H_{exch}(t)$ of 
the hole spin with the photoexcited deviation
$\Delta {\bf S}$(t) of the collective Mn spin ${\bf S}$ from ${\bf S}_{0}$
\cite{chovan,kapet}.
Here we treat $H_{exch}(t)$  at the Hartree--Fock level, 
which conserves the magnetization amplitude, and  
note that demagnetization \cite{theory} arises from correlations 
considered in Ref. \cite{kapet-corr}.  
$H_{L}(t)$  describes the interband 
dipole  coupling 
of the optical field
${\bf E}$(t),  
which propagates along
$[001]$  and is 
 linearly polarized
at an angle of $\sim$2$^{o}$ from  $[100]$.  
  $H_{L} $ 
leads to nonlinear couplings between the bands,  
 characterized by 
the Rabi energies 
$d_{nm{\bf k}}(t)$=$d_{nm{\bf k}} 
\exp[-i \omega_p t -t^2/{\tau}^{2}_{p}]$,
where 
$\omega_p$=3eV,  
$\tau_{p}$=100fs, and 
$d_{nm{\bf k}}$=$\mu_{nm{\bf k}}{\bf E}$.
The  matrix elements
$ \mu_{nm{\bf k}}$ were obtained 
from the tight-binding parameters 
of $H_b$
as in Ref.\cite{Lew}.
We derived
coupled density matrix equations of motion 
for ${\bf S}$, all nonthermal  populations, and 
the photoinduced transient coherences between {\em all} 
conduction and valence bands \cite{chovan,kapet}. 
The dynamics of these coherences,  
neglected within the semiclassical 
rate equation treatment of spin photoexcitation, 
is important for determining the photoexcited 
hole spin ${\bf s}_h$(t).
For details, see Refs.
\cite{chovan,kapet,vogl,Lew} and a future publication.  

The Mn spin  is obtained from the  
equation of motion 
\begin{equation}\label{eq:MnSpin}
\partial_{t}{\bf S}= -\beta{\bf S}\times
{\bf s}_h
-\gamma {\bf S}\times {\bf H}_{\rm th}
+\frac{\alpha}{S} {\bf S}\times \partial_{t}{\bf S},
\end{equation}
where $\gamma$ is the gyromagnetic ratio
and  $\alpha$ the Gilbert damping coefficient \cite{Qi}. 
Our mean--field approximation misses demagnetization
 \cite{theory}, 
which can be included phenomenologically \cite{bigot2}
or  microscopically \cite{kapet-corr}. 
Here, the photoexcited hole spin 
creates 
an effective  magnetic field pulse,  
$\gamma {\bf H}_{ph}$(t)=$\beta {\bf s}_h$(t), 
which transiently modifies the  magnetic anisotropy 
and triggers nonthermal switching.
${\bf s}_h$
is 
obtained microscopically 
after noting that 
\begin{equation}
\label{eq:holeSpin}
{\bf s}_{h}=\frac{1}{V}\sum_{n{\bf k}} {\bf s}^h_{{\bf k}nn}
 \langle \hat{h}^{\dag}_{{\bf k}n}\hat{h}_{{\bf k}n}\rangle
+ 
\frac{1}{V}\sum_{n \ne n'{\bf k}} {\bf s}^h_{{\bf k}nn'}
\langle \hat{h}^{\dag}_{{\bf k}n}\hat{h}_{{\bf k}n'}\rangle,
\end{equation} 
where $V$ is the volume. 
${\bf s}^h_{{\bf k}nm}$ are
the hole spin matrix elements 
with respect to the  eigenstates of $H_b$.
They describe  spin mixing  
due to  $H_{SO}$ and $H_{pd}$.
The first term in Eq.(\ref{eq:holeSpin}) is 
roughly proportional to the photoexcited hole density. 
For $\omega_p$=3eV, it comes from 
the transient population of 
high energy  states close to the $\Lambda$ point; these
do not contribute to the ground state
magnetic anisotropy. 
The 
second term in Eq.(\ref{eq:holeSpin}) 
comes from the 
 coherences $\langle \hat{h}^{\dag}_{{\bf k}n}\hat{h}_{{\bf k}n'}\rangle$, 
$n$$\ne$$n'$, 
photoinduced 
via Raman processes. 
Here we solved the nonlinear equations of motion 
that give all $\langle \hat{h}^{\dag}_{{\bf k}n}\hat{h}_{{\bf k}m}\rangle$
at the eight ${\bf k}$'s with maximum density of states 
($\Lambda$--point). 
The  mean field arising from all ${\bf k}$ 
will be considered elsewhere. 
 The dynamics 
depends on $H_{exch}(t)$   
and the nonlinear polarizations $\langle \hat{h}_{-{\bf k} n} 
\hat{e}_{{\bf k} m} \rangle$, 
obtained 
from their equations of motion \cite{chovan,kapet}. 
We thus 
describe the effect on  ${\bf s}_h$(t) 
of (i) 
static bandstructure and the competition 
of spin--orbit and magnetic exchange interactions, 
(ii) photoinduced interactions that transiently 
change the bandstructure,  
and (iii) nonlinear coherence.
We 
considered dephasing/relaxation times 
$\sim$30fs \cite{jung}.

The $\Gamma$--point {\em thermal}  hole Fermi sea 
\cite{jung}
is clearly 
distinguished  from the carriers
that create ${\bf s}_h$(t), which are 
 photoexcited 
at $\sim$3eV.
Our main assumption is that, within time intervals $\sim$100fs,
the Fermi sea 
adjusts adiabatically 
to the transient changes in 
the Mn spin order parameter and   
can thus be described 
in terms of its
total energy
$E_{h}[{\bf S}(t)]$. $E_h$ then   
depends on the instantaneous Mn spin 
(unit vector 
$\hat{{\bf S}}$) 
and the second term in 
Eq.(\ref{eq:MnSpin}) 
describes 
precession 
around the {\em thermal} magnetic 
anisotropy 
field 
$\gamma {\bf H}_{ \rm th}$=$-\partial E_{h}/\partial {\bf S}$
\cite{kimel-rev,bigot2}.  
Symmetry \cite{symmetry,welp} 
dictates that, independent of issues  
surrounding the bands 
close to the bandgap 
of (Ga,Mn)As  \cite{jung,theory},
\begin{equation}
\label{eq:Eh}
E_h=
K_c(\hat{S}^2_x \hat{S}^2_y+ \hat{S}^2_x \hat{S}^2_z
+ \hat{S}^2_y \hat{S}^2_z)+K_{uz} \hat{S}^2_z-K_{u} \hat{S}_x \hat{S}_y
- \gamma H S_z
\end{equation}
as observed experimentally \cite{welp}.  
$K_c$ is the cubic anisotropy constant, 
$K_{uz}$ is the uniaxial constant due to strain and shape
anisotropies,  $K_u$ describes an in--plane  anisotropy 
that may be due to materials issues \cite{welp}, 
and $H$ is an external magnetic field  along $[001]$.
We extract these parameters  from 
experiment \cite{welp} and neglect any 
 transient temperature effects, 
which enhance our predicted anisotropy and
give demagnetization \cite{bigot2,theory}. 
We thus obtain 
a {\em biaxial} magnetic
anisotropy with {\em four} metastable 
magnetic ground states \cite{asta,wang-apl}.
Since the easy axes are close to  
y=0 or x=0, we label these four magnetic ground states by
$X^\pm$ and $Y^\pm$ \cite{wang-apl}. 

\begin{figure}[t]
\begin{center}
\includegraphics[scale=1,angle=0]{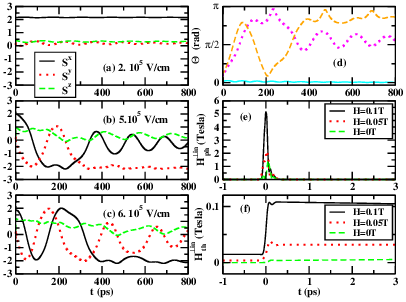}
\caption{(color online) (a)--(c): ${\bf S}$(t) 
for $H$=100mT, $\alpha$=0.03 \cite{Qi}, 
 and three pump fluences.
$K_c$=0.0144meV, $K_u$=0.0025meV, and  
$K_{uz}$=5$K_c$
give  anisotropy fields  1.7mT, 0.29mT, and 8.3 mT  \cite{jung}. 
(d): Angle 
$\cos \Theta$=${\bf \hat{S}}_{0}\cdot{\bf \hat{S}}$(t), 
0$\le$$\Theta$(t)$\le$$\pi$.  
(e): In--plane component of the  non--thermal   field  
$\gamma {\bf H}_{ph}^{\bot}$=$\beta ({\bf s}_{h})^{\bot}$. 
(f):  Thermal  field component ${\bf H}^{\bot}_{th}$.
}\label{fig1}
\end{center}
\end{figure}

Figures \ref{fig1}(a), (b) and (c) show ${\bf S}$(t)
for initial condition along $X^+$
and  
three pump fluences.  
They clearly demonstrate magnetic switching  controlled nonthermally 
by  the pump pulse. 
Fig. \ref{fig1}(a)
was obtained 
for  $E$=2$\times$10$^{5}$V/cm, as in the experiment of 
Ref.\cite{wang-apl} (fluence  $\sim$7$\mu$J/cm$^2$). 
In this case,  
$\Delta {\bf S}$ is small
and  ${\bf s}_h$ excites  magnon oscillations. These results
agree with Ref.\cite{wang-apl}. 
With increasing pump intensity, the 
magnetic  nonlinearities 
of 
Eq. (\ref{eq:Eh})
kick in.
For $E$=5$\times$10$^{5}$V/cm,  we obtain 
$X$$\to$$Y$ switching after $\sim$400ps (to $Y^-$), while 
for intermediate times, the magnetization 
reverses direction (to $X^-$, Fig.\ref{fig1} (b)).  
For
$E$=6$\times$10$^{5}$V/cm, the magnetization reversal is complete
after $\sim$400ps
(to $X^-$, Fig.\ref{fig1} (c)).
Fig.\ref{fig1}(d) shows this switching  more clearly,  
by plotting the magnetization reorientation 
angle $\Theta(t)$ = $\cos^{-1} ({\bf \hat{S}}_{0}\cdot {\bf \hat{S}})$
(0$\le$$\Theta$$\le$$\pi$).  
Our scheme 
could thus potentially be used 
to write  multiple
bits in a massively parallel memory
with  THz
speed, since one normally reads bits long after writing them.

Our 
switching 
is initiated by 
the photoinduced non-thermal (${\bf H}_{ph}$) 
field 
(Fig.\ref{fig1}(e))
and completed by the thermal (${\bf H}_{th}$) field 
(Fig. \ref{fig1}(f)).
${\bf H}_{ph}$
only lasts  for $\sim$100fs and is 
enhanced by the external 
magnetic field
$H$.
For given intensity, 
${\bf H}_{ph}$ is much stronger 
for $\omega_p$=3eV 
than for 
$\omega_p$=1.5eV 
due to the difference in the density of states 
and spin--orbit interaction \cite{kapet}.
In contrast, the thermal field 
${\bf H}_{th}$ develops 
as $E_h$ changes due to the Mn spin 
$\Delta {\bf S}$(t) photoinduced by ${\bf s}_h$. 
It is much weaker than 
${\bf H}_{ph}$ during the first 100's of fs
but dominates over $\sim$100ps.

\begin{figure}[t]
\begin{center}
\includegraphics[scale=1]{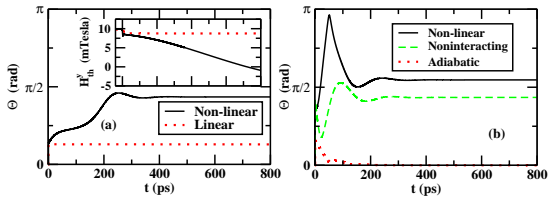}
\caption{(color online) Nonlinear magnetization
dynamics ($\alpha$=0.3, 
$E$=6$\times$10$^{5}$V/cm). 
(a): Comparison of Mn spin re--orientation angles 
$\Theta$(t) and thermal anisotropy field $H_{th}^y$ 
along [010] (inset) during  $\sim$100ps 
and 
$\sim$10ps timescales. 
(b): Comparison of full calculation (solid line)
with  result 
obtained by neglecting 
$H_{exch}$(t) 
(dashed line)
and within semiclassical adiabatic approximation of spin photoexcitation  
(dotted line).
}\label{fig2}
\end{center}
\end{figure}

We now turn to the origin  of the photo--induced nonlinear behavior
demonstrated by Fig.\ref{fig1}. 
By expanding  ${\bf H}_{th}$ 
to first order in 
$\Delta {\bf S}$, 
we obtain the linear spin dynamics shown 
by the dotted lines in Fig.\ref{fig2} (a).    
The difference between 
the linear and nonlinear Mn spin re--orientation angle 
$\Theta$(t) 
is small during fs timescales
but grows over many ps. 
To interpret this, 
the inset of Fig. \ref{fig2} shows that 
$H_{th}^{y}$, the  
component perpendicular to ${\bf S}_0$, 
grows over 
 10's of ps
due to the 
cubic anisotropy nonlinearity $\propto$$S_y^3$, 
as $S_y$ increases due to  precession  around
$H_{th}^z$. 
In turn, $S_z$ (and thus  
$H_{th}^z$) 
increases 
due to precession around 
 $H_{th}^y$.
Switching is triggered by this nonlinear dependence,  
as 
${\bf H}_{th}$ builds
over 10's of ps until the magnetization overcomes 
the energy barrier between the  magnetic states. 

Fig.\ref{fig2}(b)) shows that 
the 
time--dependent changes 
in the bandstructure  $H_b$,  
due to the photoinduced interaction $H_{exch}$(t), 
affect ${\bf s}_h$
and  $\Theta$(t) (compare solid and dashed 
lines).  
The dotted line in  Fig.\ref{fig2}(b) 
also shows that the 
 semiclassical 
Fermi's Golden rule  hole spin generation rate, 
obtained by solving our 
equations within the adiabatic approximation 
while treating the interactions  
perturbatively \cite{rossi}, 
can also give 
strong deviations. We conclude that 
our treatment of the 
coherent nonlinear photoexcitation of the interacting system is  important 
for modelling the  magnetization  dynamics.

In summary, we predict spin switchings triggered by
coherent nonlinear photoexcitation of 
(III,Mn)V ferromagnets 
by a {\em single} femtosecond optical pulse.  
We obtained 
all--optical 
magnetization reversal 
and demonstrated a four--state 
magnetic switching functionality  
 resulting from the interplay 
between  ultrafast coherence, nonlinearity, and 
competing spin interactions. Our proposed effect 
should be confirmed experimentally with femtosecond magneto--optical 
spectroscopy 
and points to future possibilities
for further reducing the switching times 
by using multiple coherent optical pulses.

This work was supported by the EU ITN program ICARUS, 
the U.S. National Science Foundation grant 
DMR-1055352,  
and the U.S. Department of Energy-Basic Energy Sciences under contract 
DE-AC02-7CH11358.

\end{document}